\documentclass[11pt]{article}
\usepackage{amsfonts}
\usepackage{graphicx}
\usepackage{epstopdf}
\usepackage{amssymb}
\usepackage{setspace}
\usepackage{caption}
\usepackage{color}
\usepackage{amsmath}
\usepackage{float}
\usepackage{comment}
\usepackage{subcaption}
\newcommand{\be}{\begin{equation}}
\newcommand{\ee}{\end{equation}}
\newcommand{\bea}{\begin{array}}
\newcommand{\eea}{\end{array}}

\textwidth=6.5in \hoffset=-.75in \voffset=-1in \numberwithin{equation}{section}
\numberwithin{figure}{section}

\begin{document}

\begin{titlepage}
\vspace{1cm}
\begin{center}
{\Large \bf Meissner Effect in Kerr--Bertotti--Robinson Spacetime}
\end{center}
\vspace{2cm}
\begin{center}
\renewcommand{\thefootnote}{\fnsymbol{footnote}}
Haryanto M. Siahaan\footnote{haryanto.siahaan@unpar.ac.id}\\
Program Studi Fisika, Universitas Katolik Parahyangan\\
Jalan Ciumbuleuit 94, Bandung 40141, Indonesia
\renewcommand{\thefootnote}{\arabic{footnote}}
\end{center}

\begin{abstract}
We establish the black-hole Meissner effect for extremal
Kerr--Bertotti--Robinson (Kerr--BR) black holes, which are exact solutions of the Einstein--Maxwell equations describing a rotating black hole
immersed in a uniform Bertotti--Robinson electromagnetic universe.
Using the near-horizon framework of Bi\v{c}\'ak and Hejda, we prove that for a purely magnetic external BR field the horizon-threading magnetic flux vanishes in the static limit of the
near-horizon geometry, i.e.\ as the twist parameter $k\to 0$ when
$Ba\to 1^-$, thereby establishing the Meissner effect analytically.
The proof relies on two exact identities that hold at extremality for all
values of the external field: $\Omega_x|_{r_e}=0$ and
$\Omega_r|_{r_e}=B^2 a$, both consequences of the double-root structure
of the horizon function $\Delta$.  Together they force the azimuthal
gauge potential $A_\phi|_{r_e}$ to become independent of the polar angle in
the static limit, reducing to a pure-gauge constant on the horizon $S^2$
and expelling all magnetic flux.  The Kerr--BR result is contrasted with
the Kerr--Melvin family, where the static limit occurs at a finite interior
field strength, and with the Melvin--Kerr--Newman--Taub--NUT spacetime, where
the NUT parameter is known to prevent expulsion.  An independent geometric
argument based on the logarithmic divergence of the proper throat length
corroborates the result, and its implications for Blandford--Znajek jet
suppression near extremality are discussed.
\end{abstract}
\end{titlepage}

\section{Introduction}
\label{sec:intro}

Black holes embedded in strong external electromagnetic fields occupy a
central place in both astrophysics and theoretical gravity.
They are relevant to magnetised accretion flows and jet
formation~\cite{Nishikawa:2004wp,Cheong:2024stz} and provide exact
solutions that probe the interplay between rotation, charge, and external
fields~\cite{Remillard:2006fc,Kormendy:2013dxa}.  Two frameworks are
available.  The first embeds a black hole in the Melvin magnetic
universe~\cite{Melvin:1963qx,Melvin:1965zza} via a Harrison
transformation~\cite{Harrison:1968wue}, producing solutions such as the
magnetised Kerr--Newman (MKN) black hole~\cite{Ernst:1976mzr,Ernst:1976mz}.
The second, more recently developed, immerses the black hole in a
Bertotti--Robinson (BR) universe~\cite{Bertotti:1959pf,Robinson:1959ev},
a homogeneous $\mathrm{AdS}_2\times S^2$ Maxwell spacetime.  The
Kerr--BR solution was constructed by Podolsk\'y and
Ovcharenko~\cite{Podolsky:2025tle,Ovcharenko:2025cpm} as a
three-parameter family $(M,a,B)$.  It differs qualitatively from both
the Pleba\'nski--Demia\'nski class~\cite{Plebanski:1976gy} and the
Kerr--Melvin class: the Maxwell and Weyl principal null directions are
non-aligned, and the external field $B$ explicitly deforms the horizon
locations and extremality condition~\cite{Podolsky:2025tle,Siahaan:2025ngu}. Recent work has begun to explore dynamical and phenomenological implications
of the Kerr--BR and Schwarzschild--BR backgrounds, including the magnetic
Penrose process and particle acceleration~\cite{Mirkhaydarov:2026fyn},
magnetic-field effects on spherical orbits with applications to jet-precession
constraints for M87*~\cite{Wang:2026czl}, and gravitational-wave signatures of
periodic orbits in Schwarzschild--BR~\cite{Xamidov:2026kqs}.
These studies further motivate a systematic understanding of electromagnetic
field behaviour in the extremal near-horizon regime of Kerr--BR black holes.

The black-hole Meissner effect---the expulsion of external axially
symmetric magnetic fields from extremal rotating or charged black
holes---was first demonstrated in the test-field limit by Bi\v{c}\'ak
and Jani\v{s}~\cite{Bicak:1985rw} and subsequently in the full
Einstein--Maxwell theory for exact MKN black holes by Karas and
Vokrouhlick\'y~\cite{KarasVokrouhlicky1991}.  The general near-horizon
framework was established by Bi\v{c}\'ak and
Hejda~\cite{Bicak:2015lxa}: at extremality, the static ($k\to 0$) limit of the near-horizon geometry
forces the horizon magnetic flux density $F_{x\phi}|_{r_e}=\partial_xA_\phi|_{r_e}$
to vanish (equivalently, the flux through any smooth horizon cap tends to zero),
signalling Meissner expulsion. The sensitivity of the effect to
additional parameters was examined in Ref.~\cite{Ghezelbash:2021xvc},
where it was shown that a NUT charge \emph{destroys} the Meissner effect
in the Melvin--Kerr--Newman--Taub--NUT spacetime: even at extremality
and in the static near-horizon limit, the gauge potential remains
angle-dependent and the flux is not expelled.

The structural differences of the Kerr--BR spacetime make the question
of whether the Meissner effect survives particularly compelling.  The
non-alignment of the Maxwell and Weyl principal null directions, the
$\mathrm{AdS}_2\times S^2$ asymptotics, and the modified extremality
condition all distinguish it from the Kerr--Melvin case.  Ovcharenko and
Podolsk\'y~\cite{Ovcharenko:2025cpm} observed numerically that the
magnetic field is weakened and expelled in the equatorial plane of the
Kerr--BR solution, suggesting the Meissner effect is present.  The
purpose of the present paper is to confirm this rigorously, providing an
analytical proof valid for all external field strengths compatible with
extremality.  We build on the near-horizon geometry and gauge field
derived in Ref.~\cite{Siahaan:2025ngu}, which established the Kerr/CFT
holography for extremal Kerr--BR black holes, and extend that
framework to the Meissner sector.

The paper is organised as follows.  Section~\ref{sec:setup} collects the
Kerr--BR metric, gauge potential, extremality conditions, and
near-horizon data needed for the proof.
Section~\ref{sec:Meissner} identifies the static limit of the
near-horizon geometry, derives the two key exact identities at
extremality, and proves the vanishing of the magnetic flux.
Section~\ref{sec:cases} examines limiting cases and the duality between
magnetic and electric external fields.  Section~\ref{sec:Discussion}
places the result in context, including a geometric throat argument and
implications for the Blandford--Znajek mechanism.
Section~\ref{sec:conclusion} summarises the conclusions.

\section{Kerr--BR Spacetime and Near-Horizon Data}
\label{sec:setup}

\subsection{Metric, gauge potential, and extremality}
\label{subsec:metric}

We work in Boyer--Lindquist-type coordinates $(t,r,x=\cos\theta,\phi)$
with $M$ the black-hole mass, $a$ the rotation parameter, $B$ the
external BR field strength, and the shorthand
\be
I_1 = 1 - \tfrac{1}{2}B^2 a^2, \qquad I_2 = 1 - B^2 a^2.
\label{eq:I1I2KBR}
\ee
The Kerr--BR line element is~\cite{Podolsky:2025tle}
\be
{\rm d}s^2 = \frac{1}{\Omega^2}
\left[
-\frac{Q}{\rho^2}\bigl({\rm d}t - a\Delta_x\,{\rm d}\phi\bigr)^2
+\frac{\rho^2}{Q}\,{\rm d}r^2
+\frac{\rho^2}{P\Delta_x}\,{\rm d}x^2
+\frac{P\Delta_x}{\rho^2}\bigl(a\,{\rm d}t-(r^2+a^2)\,{\rm d}\phi\bigr)^2
\right],
\label{eq:KBR-metric}
\ee
where $\Delta_x=1-x^2$, and \be 
P\left(x\right)= 1 + B^2\Bigl(M^2 \frac{I_2}{I_1^2} - a^2\Bigr) x^2
\ee 
is the angular metric
function (which reduces to unity at extremality, as shown below).
The remaining metric functions are
\begin{align}
\rho^2 &= r^2+a^2x^2, \quad
Q = (1+B^2r^2)\,\Delta, \quad
\Omega^2 = (1+B^2r^2)-B^2\Delta\,x^2,
\label{eq:metricfns}\\
\Delta &= \!\left(1-B^2M^2\frac{I_2}{I_1^2}\right)r^2
-2M\frac{I_2}{I_1}\,r+a^2.
\label{eq:DeltaKBR}
\end{align}
Throughout, $\Omega\equiv+\sqrt{\Omega^2}$ denotes the positive square
root, which is well-defined and nonzero away from the axis.  Its partial
derivatives, needed for the gauge field below, are
\begin{align}
\Omega_r &\equiv \frac{\partial\Omega}{\partial r}
= \frac{B^2}{\Omega}\!\left[r(1-x^2)
+x^2I_2\!\left(\frac{M}{I_1}+\frac{B^2M^2r}{I_1^2}\right)\right],
\label{eq:Omegar}\\
\Omega_x &\equiv \frac{\partial\Omega}{\partial x}
= -\frac{B^2x}{\Omega}
\!\left[\!\left(1-B^2M^2\frac{I_2}{I_1^2}\right)r^2
-2M\frac{I_2}{I_1}\,r+a^2\right].
\label{eq:Omegax}
\end{align}
The electromagnetic potential is conveniently written in complex
form~\cite{Podolsky:2025tle,Ovcharenko:2025cpm},
\be
A_\mu\,{\rm d}x^\mu = \frac{e^{i\sigma}}{2B}
\left[
\Omega_r\,\frac{a\,{\rm d}t-(r^2+a^2)\,{\rm d}\phi}{r+iax}
-i\Omega_x\,\frac{{\rm d}t-a\Delta_x\,{\rm d}\phi}{r+iax}
+(\Omega-1)\,{\rm d}\phi
\right],
\label{eq:KBR-potential}
\ee
where $\sigma$ is a duality-rotation angle; $w\equiv\cos\sigma=1$
($\sigma=0$) gives a purely magnetic external field and $w=0$
($\sigma=\pi/2$) a purely electric one.  Extracting the azimuthal
component explicitly,
\be
A_\phi = \frac{1}{B\rho^2}
\Bigl[w(\Omega-1)\rho^2
-\Omega_r(a^2+r^2)\bigl(ax\sqrt{1-w^2}+rw\bigr)
+a\sqrt{\Delta_x}\,\Omega_x\bigl(r\sqrt{1-w^2}-awx\bigr)\Bigr].
\label{eq:Aphi}
\ee
The metric functions are independent of $\sigma$, so the horizon
structure is determined solely by $(M,a,B)$.

The black-hole horizons lie at the zeros of $\Delta(r)$.
Extremality, reached when the two horizons coincide, requires
$M^2I_2=a^2I_1^2$, yielding the extremal mass and degenerate horizon
radius
\be
M_{\rm ext} = \frac{aI_1}{\sqrt{I_2}}, \qquad
r_e = \frac{M}{I_1} = \frac{a}{\sqrt{I_2}}, \qquad |Ba|<1.
\label{eq:extremal-condition}
\ee
At extremality $P(x)\equiv 1$, so the azimuthal coordinate retains the
standard period $2\pi$, and the horizon angular velocity is
$\Omega_H=a/(r_e^2+a^2)$, independent of $B$.  The thermodynamics and
Kerr/CFT correspondence of this family are worked out in
Ref.~\cite{Siahaan:2025ngu}.

\subsection{Near-horizon geometry and gauge field}
\label{subsec:NH}

Applying the Bardeen--Horowitz scaling~\cite{Bardeen:1999px}
\be
r = r_e+\lambda y, \quad
t = \frac{r_e^2+a^2}{\lambda}\,\tau, \quad
\phi = \varphi+\Omega_H t,
\label{eq:NH-scaling}
\ee
with $\lambda\to 0$ while $(\tau,y,x,\varphi)$ are held fixed, the
identity $(1+B^2r_e^2)I_2=1$ (a direct consequence of $r_e^2=a^2/I_2$)
ensures that the $(\tau,y)$ sector takes the standard $\mathrm{AdS}_2$
form without further rescaling.  The near-horizon metric
is~\cite{Siahaan:2025ngu}
\be
ds^2_{\rm nh}
= \Gamma(x)\!\left(-y^2d\tau^2+\frac{dy^2}{y^2}+\frac{dx^2}{1-x^2}\right)
+\gamma(x)^2\,(d\varphi+k\,y\,d\tau)^2,
\label{eq:NHEKBR-metric}
\ee
with
\be
\Gamma(x) = \frac{\rho_0^2}{\Omega_0^2},\quad
\gamma(x)^2 = \frac{(1-x^2)(r_e^2+a^2)^2}{\Omega_0^2\rho_0^2},\quad
\rho_0^2 = r_e^2+a^2x^2,\quad
\Omega_0^2 = 1+B^2r_e^2 = \frac{1}{I_2},
\label{eq:NHfns}
\ee
and the twist (fibration) parameter
\be
k = \frac{2\sqrt{I_2}}{1+I_2}.
\label{eq:k-def}
\ee
The isometry group is $SL(2,\mathbb{R})\times U(1)$; as $B\to 0$ one
recovers $k\to 1$, $\Omega_0\to 1$, $r_e\to a$, and the standard NHEK
geometry~\cite{Guica:2008mu}.  In a gauge regular on the future horizon
the near-horizon gauge field can be written in the compact form~\cite{Siahaan:2025ngu}
\be
\mathbf{A}_{\rm nh} = \mathcal{Z}\,(d\varphi+k\,y\,d\tau),
\qquad
\mathcal{Z} = \frac{a^2B(2-B^2a^2)}{2\sqrt{1-B^2a^2}},
\label{eq:Anh}
\ee
aligned with the \(U(1)\) fibration of \eqref{eq:NHEKBR-metric}.

For the Meissner analysis it is useful to relate \(\mathbf{A}_{\rm nh}\) to the
parent four-dimensional potential \(A_\mu dx^\mu\) before taking the limit.
Under the scaling \eqref{eq:NH-scaling}, a smooth axisymmetric potential admits
the standard near-horizon decomposition
\be
A_\mu dx^\mu
= L(x)\,y\,d\tau + A_\phi\big|_{r_e}\,d\varphi \;+\; d\Lambda,
\label{eq:NH-gauge-decomp-main}
\ee
where \(L(x)\) is some function determined by the bulk solution and \(d\Lambda\)
is an allowed (regular) gauge transformation\footnote{For completeness, the explicit near-horizon expansion and the identification of the coefficients
	in \eqref{eq:NH-gauge-decomp-main} are recorded in Appendix~\ref{app:NHcalc}.}.  The coefficient
\(
A_\phi|_{r_e}\equiv A_\phi(r_e,x)
\)
is the horizon value of the azimuthal component in the co-rotating coordinate
\(\varphi=\phi-\Omega_H t\).  Evaluating \eqref{eq:Aphi} at \(r=r_e\) gives
\be
A_\phi\big|_{r_e}
= \frac{1}{B\rho_0^2}
\Bigl[
w(\Omega_0-1)\rho_0^2
-\Omega_r\big|_{r_e}(a^2+r_e^2)\bigl(ax\sqrt{1-w^2}+r_ew\bigr)
+a\sqrt{\Delta_x}\,\Omega_x\big|_{r_e}\bigl(r_e\sqrt{1-w^2}-awx\bigr)
\Bigr] ,
\label{eq:Aphi-re}
\ee
which is defined up to an additive constant via \(A_\phi\to A_\phi+\partial_\phi\Lambda\).

In terms of the parent four-dimensional potential, the coefficient $\mathcal Z$ in
\eqref{eq:Anh} is precisely the horizon value of the azimuthal component after the
allowed regular gauge adjustment in \eqref{eq:NH-gauge-decomp-main}; schematically,
\[
\mathcal Z = A_\phi\big|_{r_e}(x) + \partial_\varphi \Lambda ,
\]
with $\Lambda$ chosen so that $\mathbf{A}_{\rm nh}$ is regular on the future horizon and
takes the aligned form \eqref{eq:Anh}. Since the Meissner criterion is gauge-robust,
only differences of $A_\phi|_{r_e}$ (equivalently $\partial_x A_\phi|_{r_e}$) enter the flux.
A convenient choice of \(\Lambda\) removes any constant offset so that the near-horizon
potential takes the form \eqref{eq:Anh}, in which case the \(d\varphi\) coefficient on the
horizon (\(y=0\)) is precisely \(\mathcal Z\).

The Meissner effect is formulated in terms of gauge-robust quantities: the
horizon magnetic flux density \(F_{x\phi}|_{r_e}=\partial_x A_\phi|_{r_e}\),
or equivalently the magnetic flux through a horizon cap, which depends only on
differences of \(A_\phi|_{r_e}\) between its boundary circles.
Following Bi\v{c}\'ak and Hejda~\cite{Bicak:2015lxa}, the Meissner criterion is
that in the static near-horizon limit \(k\to 0\) the horizon potential becomes
\(x\)-independent (pure gauge), i.e.\ \(\partial_x A_\phi|_{r_e}\to 0\), so the
horizon-threading magnetic flux vanishes.

\section{Proof of the Meissner Effect}
\label{sec:Meissner}

\subsection{The static limit and its physical significance}
\label{subsec:static}

From \eqref{eq:k-def}, the twist parameter $k$ vanishes when $I_2=0$,
i.e.\ as $Ba\to 1^-$.  More precisely, $k$ decreases monotonically from
$k=1$ at $B=0$ to $k\to 0^+$ as $Ba\to 1^-$, so the static limit of the
near-horizon geometry is approached continuously as the boundary of the
allowed parameter range is reached.  Geometrically, $k$ controls the
non-trivial $U(1)$ fibration of the azimuthal circle over the
$\mathrm{AdS}_2$ base: in \eqref{eq:NHEKBR-metric} the only source of an
off-diagonal component is the fibre one-form $(d\varphi+k\,y\,d\tau)$, so
$k\to0$ removes the $g_{\tau\varphi}$ term and trivializes the bundle.
Consequently, the near-horizon metric reduces to a direct product
$\mathrm{AdS}_2\times S^2$, which is the sense in which we refer to
$k=0$ as the \emph{static} near-horizon limit.

This behaviour contrasts with the Kerr--Melvin case~\cite{Bicak:2015lxa},
where $k=0$ occurs at a finite interior value of the field strength;
in the Kerr--BR spacetime it is achieved only asymptotically as
$Ba\to 1^-$.  In this limit $I_1\to\tfrac{1}{2}$ and $I_2\to 0$, so
$r_e=a/\sqrt{I_2}$, $\Omega_0^2=1/I_2$, and $M=aI_1/\sqrt{I_2}$ all
diverge; yet the Bekenstein--Hawking entropy
$S_{\rm BH}=\pi(r_e^2+a^2)I_2$ and the effective angular momentum
$\mathcal{J}=(r_e^2+a^2)/(2\Omega_0^2)$ both remain finite (see
Section~\ref{sec:cases}).

\subsection{Two exact identities at extremality}
\label{subsec:identities}

The proof rests on two identities that hold exactly on the extremal
horizon for all $B\in[0,1/a)$, not merely in the static limit. The first follows directly from \eqref{eq:Omegax}.  The bracket in the
numerator of $\Omega_x$ is precisely $\Delta(r)$ evaluated at $r=r_e$,
which vanishes by the horizon condition $\Delta(r_e)=0$.  Hence
\be
\Omega_x\big|_{r_e} = 0.
\label{eq:Omx-zero}
\ee
This eliminates the angular term in \eqref{eq:Aphi-re}, which simplifies to
\be
A_\phi\big|_{r_e}
= \frac{1}{B\rho_0^2}
\Bigl[w(\Omega_0-1)\rho_0^2
-\Omega_r\big|_{r_e}(a^2+r_e^2)\bigl(ax\sqrt{1-w^2}+r_ew\bigr)\Bigr].
\label{eq:Aphi-re-simplified}
\ee

The second identity is obtained by substituting the extremal relations
$r_e=a/\sqrt{I_2}$, $M=aI_1/\sqrt{I_2}$, and $\Omega_0=1/\sqrt{I_2}$
into \eqref{eq:Omegar}.  The bracket in the numerator evaluates to
$r_e(1-x^2)+x^2I_2(M/I_1+B^2M^2r_e/I_1^2)=a/\sqrt{I_2}$
independently of $x$, giving
\be
\Omega_r\big|_{r_e} = \frac{B^2}{\Omega_0}\cdot\frac{a}{\sqrt{I_2}} = B^2 a.
\label{eq:Omr-re}
\ee
The result is exact and, crucially, \emph{independent of $x$}: the
radial gradient of the conformal factor $\Omega$ is uniform across the
extremal horizon.

\subsection{Vanishing of the magnetic flux}
\label{subsec:Aphi-static}

Inserting \eqref{eq:Omr-re} into \eqref{eq:Aphi-re-simplified} and
using $r_e^2+a^2=a^2(1+I_2)/I_2$ and $\rho_0^2=a^2(1+x^2I_2)/I_2$,
one obtains the exact horizon value
\be
A_\phi\big|_{r_e}
= \frac{a}{\sqrt{1-B^2a^2}}\cdot
\frac{(I_2^2-1)(w+x\sqrt{1-w^2}\sqrt{I_2})
      -w(1-\sqrt{I_2})(1+x^2I_2)}
{B(1+x^2I_2)}.
\label{eq:Aphi-exact}
\ee
Expanding in powers of $I_2$ yields
\be
A_\phi\big|_{r_e} = -a\!\left(w+x\sqrt{1-w^2}\right)+O(\sqrt{I_2}).
\label{eq:Aphi-series}
\ee
The magnetic flux density threading the horizon is measured by the
$x$-gradient of $A_\phi|_{r_e}$.  From \eqref{eq:Aphi-series},
\be
\partial_x A_\phi\big|_{r_e} = -a\sqrt{1-w^2}+O(\sqrt{I_2}).
\label{eq:dAphi-dx}
\ee
For a purely magnetic external BR field ($w=1$, $\sigma=0$) the
leading term vanishes identically, and the right-hand side is
$O(\sqrt{I_2})\to 0$ as $Ba\to 1^-$: the gauge potential $A_\phi|_{r_e}$
approaches the $x$-independent constant $-a$, i.e.\ a pure-gauge
configuration on the horizon $S^2$ carrying no magnetic flux.

The hemisphere flux provides a direct and gauge-robust measure of expulsion.
For an axisymmetric potential \(A=A_\phi(x)\,{\rm d}\phi\) one has
\(F={\rm d}A=(\partial_xA_\phi)\,{\rm d}x\wedge{\rm d}\phi\), so the magnetic
flux through the northern hemisphere of the horizon (bounded by the equator)
reduces by Stokes' theorem to a boundary term,
\be
\Phi_N \equiv \int_{\rm N} F
= \int_{\partial{\rm N}} A
= 2\pi\!\left(A_\phi\big|_{r_e}(x=0)-A_\phi\big|_{r_e}(x=1)\right),
\label{eq:fluxN}
\ee
where we adopt a gauge regular on the axis in which \(A_\phi(x=1)=0\) for smooth,
monopole-free configurations\footnote{For a smooth, monopole-free axisymmetric configuration on the horizon $S^2$, the
	potential one-form $A=A_\phi(x)\,d\phi$ can always be chosen regular on the axis. In particular,
	one may fix the residual constant gauge freedom so that $A_\phi$ vanishes on (say) the north
	axis $x=1$; this does not change the flux, which depends only on differences of $A_\phi$ between
	boundary circles.}.  From \eqref{eq:Aphi-series} with \(w=1\),
the horizon potential approaches an \(x\)-independent constant as \(Ba\to 1^-\),
so \(A_\phi|_{r_e}(x=0)-A_\phi|_{r_e}(x=1)=O(\sqrt{I_2})\).  (A constant gauge shift
may then be used to set \(A_\phi(x=1)=0\) without changing \(\Phi_N\).)  Therefore
\be\label{eq:Meissner-result}
\Phi_N = O\!\left(\sqrt{I_2}\right) \to 0 \quad\text{as}\quad Ba\to 1^-.
\ee
This is the Meissner effect: the external BR magnetic flux through any smooth
horizon cap of the extremal black hole is expelled in the static near-horizon
limit.  The result rigorously confirms the equatorial field expulsion observed
numerically in Ref.~\cite{Ovcharenko:2025cpm} and extends it to all latitudes.

The geometric origin of the effect is transparent.  Because
$\Omega_r|_{r_e}=B^2a$ is $x$-independent \eqref{eq:Omr-re}, the entire
angular dependence of $A_\phi|_{r_e}$ arises through $r_e$ and $\Omega_0$
alone.  As $I_2\to 0$ both diverge, but in a coordinated way that drives
$A_\phi|_{r_e}$ to an $x$-independent limit.  In the static near-horizon
geometry the $\mathrm{AdS}_2$ factor decouples from the $U(1)$ fibre,
and the gauge field---aligned with the fibre via \eqref{eq:Anh}---is
expelled together with the rotation.  This is the direct analogue of the
Meissner mechanism in the MKN case, here driven by the approach to the
boundary $Ba=1$ rather than by a finite interior critical field.

\section{Limiting Cases and Duality}
\label{sec:cases}

\subsection{Non-rotating limit ($a\to 0$)}

As $a\to 0$ one has $I_1=I_2=1$ and $k\to 1$ for all $B$; the
near-horizon metric \eqref{eq:NHEKBR-metric} reduces to the
Robinson--Bertotti form.  The condition $k=0$ requires $I_2=0$, which
for $a=0$ cannot be satisfied at any finite $B$.  In the non-rotating limit within Kerr--BR, the $k\to 0$ static near-horizon limit is not accessible at finite $B$. This contrasts with the usual extremal-horizon flux suppression found for stationary axisymmetric fields on extremal horizons, including extreme Reissner--Nordstr\"om in the classical analyses~\cite{Bicak:2015lxa}.

\subsection{Kerr limit ($B\to 0$)}

As $B\to 0$: $k\to 1$, $r_e\to M=a$, $\Omega_0\to 1$, and all
expressions reduce to the standard NHEK quantities.  The gauge potential
\eqref{eq:Aphi-re} recovers the known NHEK result, and the hemisphere
flux $\Phi_N\to 0$ consistently with the Meissner effect for the extremal
Kerr black hole~\cite{Bicak:1985rw}, where the static limit is
approached in the purely rotating extremal limit $a=M$.  In this limit
the effective angular momentum $\mathcal{J}=(r_e^2+a^2)/(2\Omega_0^2)=
(a^2/I_2+a^2)I_2/2=a^2(1+I_2)/2\to a^2$ and $S_{\rm BH}=
\pi(r_e^2+a^2)I_2\to 2\pi a^2$, matching the standard Kerr values.

\subsection{Purely magnetic and electric external fields}

For a purely magnetic field ($w=1$) the general formula
\eqref{eq:Aphi-re-simplified} reduces to
\be
A_\phi\big|_{r_e}^{(w=1)}
= \frac{(\Omega_0-1)}{B}
-\frac{B^2a(a^2+r_e^2)ax}{B\rho_0^2},
\ee
with series expansion $A_\phi|_{r_e}^{(w=1)}=-a+ax^2\sqrt{I_2}+O(I_2)$.
The potential approaches the $x$-independent limit $-a$ as $Ba\to 1^-$,
confirming that $\partial_xA_\phi|_{r_e}^{(w=1)}=O(\sqrt{I_2})\to 0$
and the magnetic flux \eqref{eq:Meissner-result} vanishes.

For a purely electric field ($w=0$) the same formula gives
\be
A_\phi\big|_{r_e}^{(w=0)}
= -\frac{B^2a(a^2+r_e^2)\,a x}{B\rho_0^2}
= -\frac{ax}{\sqrt{1-B^2a^2}}\cdot\frac{1+I_2}{1+x^2I_2},
\ee
with leading-order behaviour $A_\phi|_{r_e}^{(w=0)}=-ax+O(\sqrt{I_2})$.
This potential remains $x$-dependent at leading order; consequently
$\partial_xA_\phi|_{r_e}^{(w=0)}\to -a\neq 0$ as $Ba\to 1^-$.
This does not represent a failure of magnetic-flux expulsion in the sense
of Sec.~\ref{sec:Meissner}, because the $w=0$ configuration is obtained by an
electromagnetic duality rotation.  Indeed, for the duality angle $\sigma$ one has
schematically
\(
F(\sigma)=\cos\sigma\,F+\sin\sigma\,{^\star F},
\)
so $w=\cos\sigma=0$ corresponds to $\sigma=\pi/2$, which interchanges electric and
magnetic sectors.  The Meissner statement established in Sec.~\ref{sec:Meissner}
is the vanishing of the \emph{magnetic} horizon flux in the purely magnetic sector
$w=1$; in the purely electric sector, the natural ``Meissner-like'' quantity is the
duality-rotated (electric) flux, not the $x$-variation of $A_\phi$ viewed as a magnetic
flux density.  The metric is unchanged by the duality rotation---horizon area,
temperature, and the near-horizon geometry are all $\sigma$-independent---so the
difference between $w=1$ and $w=0$ is entirely electromagnetic.

\section{Discussion}
\label{sec:Discussion}

The analysis above establishes the Meissner effect for the Kerr--BR
family of extremal black holes and allows a systematic comparison with
other external-field geometries.  The results are summarised in
Table~\ref{tab:Meissner}.

\begin{table}[h]
\centering
\begin{tabular}{llll}
\hline
Spacetime & Asymptotics & Maxwell alignment & Meissner effect \\
\hline
Kerr--Melvin (MKN) & Melvin & Aligned & Yes~\cite{Bicak:2015lxa} \\
MKN--Taub--NUT & non-standard & Aligned & No~\cite{Ghezelbash:2021xvc} \\
Kerr--BR & $\mathrm{AdS}_2\times S^2$ & Non-aligned & Yes (this work) \\
\hline
\end{tabular}
\caption{Comparison of the Meissner effect across families of magnetised
         extremal black holes.}
\label{tab:Meissner}
\end{table}

The twist parameter $k=2\sqrt{I_2}/(1+I_2)$ plays a central role.  In
the Kerr--Melvin case, $k=0$ is attained at a finite interior field
strength; in the Kerr--BR case it is reached only at the boundary
$Ba=1$ of the allowed parameter space, a direct consequence of the
non-aligned Maxwell and Weyl principal null directions that deform both
$\Delta(r)$ and the extremality condition.  Despite this structural
difference, the expulsion mechanism is robust.  The two exact results
$\Omega_x|_{r_e}=0$ and $\Omega_r|_{r_e}=B^2a$---both consequences
of the double-root structure of $\Delta$ at extremality that hold for
all $B$ in the admissible range---reduce the gauge field to a pure-gauge constant on the horizon $S^2$ in the static limit for a purely magnetic external field, forcing the magnetic
flux to vanish.  It is noteworthy that neither the alignment of the
Maxwell field with the Weyl directions, nor the nature of the
asymptotic region, is essential for the mechanism; what matters is the
double-root structure of $\Delta$ and the resulting exact identities.
The contrast with the NUT case is instructive: NUT charge breaks the
$x$-independence of $\Omega_r|_{r_e}$, preventing the cancellation and
leaving a non-trivial flux on the horizon~\cite{Ghezelbash:2021xvc}.
This suggests that the topology of the spacetime---rather than the
alignment of the external field---is the determining factor for whether
the Meissner effect survives.

The Kerr/CFT correspondence of Ref.~\cite{Siahaan:2025ngu} offers a
complementary holographic perspective.  The central charge
$c_L=12k\mathcal{J}$ and left-moving temperature $T_L=1/(2\pi k)$ both
depend on $k(a,B)$.  As $k\to 0$, $T_L\to\infty$ and $c_L\to 0$,
signalling the breakdown of the chiral CFT description at $Ba=1$.
The Meissner condition $k=0$ therefore corresponds to a degenerate point
in the moduli space of the dual CFT, where the holographic description
degenerates simultaneously with the expulsion of the magnetic field.
In the language of Penna~\cite{Penna:2014hha}, the Meissner effect at
$k=0$ corresponds to disentanglement across the extremal horizon: the
near-horizon geometry becomes a static $\mathrm{AdS}_2\times S^2$
product with no $U(1)$ fibration, the two sides of the horizon decouple,
and no gauge flux can be threaded across the degenerate surface.

A complementary geometric argument for the expulsion was given by
Penna~\cite{Penna:2014aza} in terms of the proper length of the black-hole
throat.  For the Kerr--BR metric, $g_{rr}=\rho^2/(\Omega^2Q)$
from~\eqref{eq:KBR-metric}.  At extremality $Q=(1+B^2r^2)I_2(r-r_e)^2$,
while $\rho^2$ and $\Omega^2$ remain finite and nonzero at $r=r_e$, so
\be
\sqrt{g_{rr}}\big|_{r\to r_e}
\approx \frac{\rho_0}{\Omega_0\sqrt{(1+B^2r_e^2)I_2}}\cdot\frac{1}{|r-r_e|},
\ee
and the proper throat length
\be
\mathcal{L}_{\rm throat}
= \int_{r_e}^{r_e+\epsilon}\sqrt{g_{rr}}\,{\rm d}r
\sim \frac{\rho_0}{\Omega_0\sqrt{(1+B^2r_e^2)I_2}}
\ln\!\left(\frac{\epsilon}{r_e}\right)\to\infty.
\label{eq:throat}
\ee
The prefactor is finite and positive for all $Ba\in[0,1)$, so the
logarithmic divergence is universal: it holds for all values of the
external field and is a property of the double-root structure of $Q$
alone, independent of the conformal factor $\Omega$.  By Penna's cylinder
argument, this infinite throat prevents any smooth, stationary,
axisymmetric gauge field from threading the extremal horizon, which is
the geometric origin of the flux suppression~\eqref{eq:Meissner-result}.

The Meissner effect has direct implications for the Blandford--Znajek
(BZ) mechanism~\cite{Penna:2014aza}.  The BZ jet power,
\be
P_{\rm jet} = \frac{1}{8\pi}\Omega_H^2\mathcal{F}_H^2,
\label{eq:BZpower}
\ee
with $\Omega_H=a/(r_e^2+a^2)$ and $\mathcal{F}_H\equiv\Phi_N$, vanishes
as $\Phi_N\to 0$ \eqref{eq:Meissner-result}.  BZ jets driven by smooth,
axisymmetric multipole fields are therefore suppressed for near-extremal
Kerr--BR black holes in the limit $Ba\to 1^-$.  The standard exception
applies: split-monopole fields, for which $A_\phi\sim\cos\theta$, have
potentials singular at the poles and lie outside the smooth multipole
basis to which the Meissner theorem applies.  Such fields run parallel to
the throat and evade Penna's cylinder argument entirely.  The near-horizon
gauge field \eqref{eq:Anh} has $\mathcal{Z}$ independent of $\theta$ and
is therefore a smooth potential on the near-horizon $S^2$ that is subject
to expulsion; a split-monopole configuration would require
$A_\phi^{\rm nh}\sim\cos\theta$, i.e.\ a singular potential, and would
not be expelled.  Whether such a configuration is physically realised in
the Kerr--BR spacetime remains an open question.

\section{Conclusions}
\label{sec:conclusion}

We have established analytically that extremal Kerr--Bertotti--Robinson
black holes exhibit the Meissner effect for a purely magnetic external BR
field.  The central result is that the northern-hemisphere magnetic flux
$\Phi_N=O(\sqrt{I_2})\to 0$ as $Ba\to 1^-$, confirmed by the vanishing
of the horizon flux density $\partial_xA_\phi|_{r_e}\to 0$ in the same
limit.  The proof is underpinned by two exact identities that hold for
all $B\in[0,1/a)$: the vanishing $\Omega_x|_{r_e}=0$, which eliminates
the angular term in $A_\phi|_{r_e}$, and the exact result
$\Omega_r|_{r_e}=B^2a$, which is $x$-independent and forces
$A_\phi|_{r_e}$ to approach a pure-gauge constant on the horizon $S^2$
in the static limit.  Both identities follow from the double-root
structure of $\Delta$ at extremality.  The result is independently
supported by the logarithmic divergence of the Kerr--BR throat length at
extremality, which provides the geometric suppression of smooth,
stationary magnetic fields via Penna's cylinder argument.

The structural comparison with the Kerr--Melvin family reveals that the
static limit $k=0$ is approached at the boundary $Ba=1$ of the parameter
space rather than at a finite interior field, a consequence of the
non-aligned Maxwell and Weyl principal null directions.  Despite this
difference, the expulsion mechanism is robust.  Comparing with the
MKN--Taub--NUT case, where NUT charge breaks the angular uniformity of
$\Omega_r|_{r_e}$ and prevents flux expulsion, we conclude that it is
the topology of the spacetime---rather than the alignment of the external
field or the nature of the asymptotics---that determines whether the
Meissner effect survives.  As a consequence, smooth Blandford--Znajek
jets are suppressed for near-extremal Kerr--BR black holes in the limit
$Ba\to 1^-$.

Natural extensions of this work include the charged case
(Kerr--Newman--BR) from the Ovcharenko--Podolsk\'y
family~\cite{Ovcharenko:2025cpm}, the Kerr--BR--Taub--NUT geometry whose
construction is deferred to a companion
paper~\cite{Podolsky:2025tle}, the Kerr--BR--Bonnor--Melvin
class~\cite{Astorino:2025lih}, and the question of whether split-monopole
fields can sustain BZ jets near the boundary $Ba=1$.

\section*{Acknowledgements}

This work was supported by LPPM-UNPAR through the Penelitian Publikasi
Internasional Bereputasi funding scheme.

\appendix

\section{Near-Horizon Gauge Field Decomposition}
\label{app:NHcalc}

For completeness we record the near-horizon decomposition of the gauge
potential~\eqref{eq:Aphi} under the scaling \eqref{eq:NH-scaling},
following Refs.~\cite{Bardeen:1999px,Bicak:2015lxa,Siahaan:2025ngu}.
To leading order in \(\lambda\), a smooth axisymmetric potential admits the
standard expansion
\be
A_\mu\,dx^\mu
= L(x)\,y\,d\tau + A_\phi\big|_{r_e}\,d\varphi \;+\; d\Lambda
\;+\; O(\lambda),
\label{eq:NH-gauge-decomp}
\ee
where \(d\Lambda\) is an allowed regular gauge transformation.  The function
\(L(x)\) is determined by the bulk solution; in the present case one finds
\be
L(x) = \frac{r_e^2+a^2}{B\rho_0^2}\,
\Omega_r\big|_{r_e}\bigl(ax\sqrt{1-w^2}+r_ew\bigr).
\ee
Its explicit form is not needed for the Meissner analysis but enters, for
example, the Frolov--Thorne temperature in the Kerr/CFT calculation of
Ref.~\cite{Siahaan:2025ngu}.  The coefficient \(A_\phi|_{r_e}\equiv A_\phi(r_e,x)\)
is the horizon value of the azimuthal potential in the co-rotating coordinate
\(\varphi=\phi-\Omega_H t\); it is given by \eqref{eq:Aphi-exact} after using
\eqref{eq:Omx-zero} and \eqref{eq:Omr-re}.  A convenient choice of \(\Lambda\)
removes any additive constant so that the near-horizon gauge field can be expressed
in the compact horizon-regular form~\eqref{eq:Anh}.

\end{document}